\documentclass[english,10.5pt,aps,prd,a4paper,preprintnumbers,floatfix,nofootinbib,showpacs,superscriptaddress, notitlepage]{revtex4-1} 
 \pdfoutput=1
\usepackage{color}
\usepackage{graphicx}
\usepackage{amsmath}
\usepackage{amssymb}
\usepackage{csquotes}
\usepackage{multirow}
\usepackage{mathtools}
\usepackage[colorlinks=true,citecolor=darkred,urlcolor=darkred, pdfborder={0 0 0}]{hyperref}
\usepackage[normalem]{ulem}
\definecolor{darkred}{rgb}{0,0.6,0}
\newcommand{\ba}{\begin{array}}
\newcommand{\ea}{\end{array}}
\def\be{\begin{equation}}
\def\ee{\end{equation}}
\def\bea{\begin{eqnarray}}
\def\eea{\end{eqnarray}}
\def\gsim{\ \rlap{\raise 2pt\hbox{$>$}}{\lower 2pt \hbox{$\sim$}}\ }
\def\lsim{\ \rlap{\raise 2pt\hbox{$<$}}{\lower 2pt \hbox{$\sim$}}\ }
\def\dslash{\kern-4pt \not{\hbox{\kern-2pt $\partial$}}}
\def\pslash{\not{\hbox{\kern-2pt p}}}

\begin{document}
\DeclareGraphicsExtensions{.eps,.ps}


\title{{$ \mu-\tau  $ Reflection Symmetry and Its Explicit Breaking for Leptogenesis  in a Minimal Seesaw Model}}



\author{Newton Nath}
\email[Email Address: ]{newton@ihep.ac.cn}
\affiliation{
Institute of High Energy Physics, Chinese Academy of Sciences, Beijing, 100049, China}
\affiliation{
School of Physical Sciences, University of Chinese Academy of Sciences, Beijing, 100049, China}


\begin{abstract}
{\noindent
The minimal seesaw framework, embroiling the Dirac neutrino mass matrix $ M_D $ and the Majorana neutrino mass matrix $ M_R $, is  quite successful to explain the current global-fit results of neutrino oscillation data. In this context, we consider most predictive forms of $ M_D $ and  $ M_R $ with two simple parameters, respectively.
 Considering these matrices, we obtain the low energy neutrino mass matrix under  type-I  seesaw formalism which obeys $ \mu-\tau $ reflection symmetry and predicts $ \theta_{23} = \pi/4$ and $ \delta = \pm \pi/2 $. In the given set-up, we also evaluate the Baryon Asymmetry of the Universe (BAU) through successful leptogenesis and find that perturbation of $ \mathcal{O}(10^{-2}) $  leads to the observed BAU and breaks exactness of the symmetry. Moreover, we also perform various correlation studies among different parameters in the  framework of broken symmetry.
 }
\end{abstract}

\maketitle

\section{Introduction}\label{sec:Intro}
In spite of its remarkable success, the Standard Model (SM) of particle physics fails to address the non-zero nature of neutrino mass. On the other hand, a large variety of neutrino oscillation experiments over past two decades have established the fact that neutrinos possess  non-zero mass and  their different flavors are substantially mixed~\cite{Patrignani:2016xqp}.
On the theoretical perspective, one of the simplest ways to describe the origin of neutrino mass is to add at least  two right-handed neutrino fields,  $N_{\mu R}, N_{\tau R} $, into the SM \cite{Guo:2006qa}.
The SM gauge invariant Lagrangian containing the neutrino Yukawa matrix and the Majorana neutrino mass matrix in the diagonal basis of charged-lepton Yukawa matrix is given by
\begin{equation}\label{eq:lag}
-\mathcal{L}\supset  \overline{L}_{\alpha L}~Y_\nu^{} N_R \widetilde{H} + \dfrac{1}{2}N^{T}_{R}C M_R N_R + \mathrm{h.c.} \; ,
\end{equation}
where $L_{\alpha L}^{} $ (for $ \alpha = e, \mu, \tau $) is the left-handed lepton doublet, $Y_\nu^{}$ represents the neutrino Yukawa matrix and $\widetilde{H} = i \sigma_2^{} H^*$ with $H$ being the Higgs doublet. Here, $ M_R $ is the right-handed Majorana neutrino mass matrix with  $N_R = (N_{\mu }, N_{\tau } )_{R}^{T}$ defined in the flavor basis and $ C $ denotes the charge-conjugation matrix. Once the Higgs field attains its  vacuum expectation value ($ vev $), i.e., $v = \langle H \rangle \approx 174~\mathrm{GeV}$ ~\cite{Patrignani:2016xqp}, one obtains the Dirac neutrino  mass term as $\overline{\nu}_{\alpha L}^{} M_D^{} N_R^{} + \mathrm{h.c.}$, where $M_D^{} = v Y_\nu^{}$ is the  Dirac neutrino mass matrix.  
In the type-I seesaw formalism~\cite{Minkowski:1977sc,Yanagida:1979as,GellMann:1980vs,Mohapatra:1979ia,Schechter:1980gr} after integrating out heavy right-handed neutrino fields one finds light neutrino mass matrix as, $  M_{\nu}^{} \approx - M_D^{} M^{-1}_R M^{T}_D $ and diagonalization of $ M_{\nu}^{}  $ leads to three active neutrino masses, $ m_1, m_2, m_3 $ \footnote{Note that in the minimal seesaw formalism lightest neutrino is always massless.}. 

Besides the theoretical origin of neutrino mass, neutrino flavor mixing is another mystifying issue which is yet unknown.
In that context, symmetry based studies have been quite successful to explain the observed mixing pattern of neutrinos ~\cite{Altarelli:2010gt,Altarelli:2012ss,Smirnov:2011jv,Ishimori:2010au,King:2013eh}. In recent years,   $\mu-\tau$ flavor symmetry advances as a well nurtured theory to explain 
flavor mixing of neutrinos which anticipates $ |V_{\mu i }| = |V_{\tau i }|$, (for i = 1, 2, 3) where $ V $ is the Pontecorvo-Maki-Nakagawa-Sakata (PMNS) matrix. 
 This symmetry predicts the maximal $2-3$ mixing angle, $\theta_{23}^{} = \pi/4$ and the maximal Dirac CP-phase, $\delta = \pm \pi/2$ which was originally proposed in Ref.~\cite{Harrison:2002et} (for recent review see Ref.~\cite{Xing:2015fdg} and the references therein).
%
Moreover, the concerned symmetry restricts two Majorana phases to $ \rho, \sigma = 0, \pi/2 $ (for details see Ref.~\cite{Grimus:2003yn,Zhou:2014sya,Xing:2015fdg,Nath:2018hjx}).
%
Because of these interesting consequences which are also compatible with  the global-fit results of neutrino oscillation data \cite{Capozzi:2016rtj,Esteban:2016qun,deSalas:2017kay}, $\mu-\tau$ reflection symmetry recently attracts a lot of attention~\cite{Ge:2010js,Ferreira:2012ri, Grimus:2012hu, Mohapatra:2012tb, Ma:2015gka,Ma:2015pma,Ma:2015fpa,He:2015xha, Joshipura:2015dsa,
Joshipura:2015zla,Chen:2015siy, Joshipura:2016hvn, Nishi:2016wki,Kitabayashi:2016zec,
Zhao:2017yvw,Rodejohann:2017lre,Liu:2017frs,Xing:2017mkx,Xing:2017cwb,Nath:2018hjx,Chakraborty:2018dew,Nath:2018xkz,Nath:2018zoi}. 

Furthermore, formulation of Baryon Asymmetry of the Universe (BAU)  through successful leptogenesis (which was originally outlined in Ref.~\cite{Fukugita:1986hr}) is another finest issue which also need to be understood yet. Interestingly, it has been observed in the class of minimal seesaw models that the CP violating phase at the low energy can be related to BAU through successful leptogenesis \cite{Endoh:2002wm,Branco:2002xf,Pascoli:2006ci}. This may provide  one-to-one correspondence between the lepton asymmetry of the universe with the CP violating phase in the leptonic sector. 
%

In this study, we consider a modified Dirac neutrino mass matrix $ M_D $ as conjectured in the  littlest seesaw model  \cite{Antusch:2011ic,King:2013xba,King:2013iva,Bjorkeroth:2014vha,King:2015aea,King:2015dvf,Bjorkeroth:2015tsa,King:2016yvg,King:2016yef,King:2018kka,King:2018fqh,Ding:2018fyz,King:2019tbt} along with a non-diagonal Majorana neutrino mass matrix $ M_R $. The essential purpose of considering such a mass model is manifold. One of the noticeably important point about  this model is that it is based on a minimum number of independent parameters and predicts normal mass ordering (NO) among the active neutrino masses (i.e., $ m_3 > m_2 > m_1  $) 
\footnote{Note that there exists another possible neutrino mass pattern of the  form  $ m_2 \approx m_1 > m_3  $ which is known as inverted mass ordering (IO). Recent results of  neutrino oscillation experiment, NO$ \nu $A~\cite{NOVA2018} disfavors ( IO, $ \delta = \pi/2$) with greater than 3$ \sigma $.}.  Moreover, the predictions of the concerned symmetry  are  in well agreement with the current global-fit results of neutrino oscillation data \cite{Capozzi:2016rtj,Esteban:2016qun,deSalas:2017kay}. Also, the ongoing neutrino oscillation experiments like NO$ \nu $A~\cite{NOVA2018} and T2K \cite{Abe:2017uxa} show  a very good agreement with this result. 
Based on the given model structure, we advance to estimate the BAU via leptogenesis in our study. However, we find that only the explicit breaking of such an exact symmetry is able to explain the observed BAU. Later, considering this breaking pattern we proceed to discuss different correlations among various neutrino oscillation parameters as well as model parameters.


Outline of this work is as follows : in next section (\ref{sec:Frame}), we present a detailed description of the concerned model which is followed by $\mu-\tau$ reflection symmetry and its predictions. Section~(\ref{sec:Leptogenesis}) is devoted to a detailed study of leptogenesis where we illustrate the importance of the explicit symmetry breaking to achieve successful leptogenesis. In section (\ref{sec:SymmetryBreaking}), we perform various correlation studies considering the breaking pattern  of the  $\mu-\tau$ reflection symmetry. Finally, we summarize our main results in section(\ref{sec:Conclusion}). In Appendix (\ref{sec:appendix}), we show $\mu-\tau$ reflection symmetry in $ M_{\nu} $ considering a different texture of  $ M_D $.
\section{General Framework}\label{sec:Frame}
In this section, we present a general formalism of $\mu-\tau$ reflection symmetry considering minimal  seesaw framework. To show that the low energy neutrino mass matrix satisfies $\mu-\tau$ reflection symmetry, we consider the field transformation as 
\begin{equation}
\nu_{eL}^{} \leftrightarrow \nu^{c}_{e L}, ~~~ \nu_{\mu L}^{} \leftrightarrow \kappa \nu^{c}_{\tau L}.
\end{equation}
 Such transformations  lead to the low energy  light neutrino mass matrix $M_\nu^{}$ as
\begin{eqnarray}\label{eq:low_mnu}
 M_\nu =  \left( \begin{matrix}
 m_{ee} &  m_{e \mu} &  \kappa m_{e \mu}^{*} \cr
\star  & m_{\mu \mu}  & m_{\mu \tau} \cr  
\star &  \star & m_{\mu \mu}^{*}  
\end{matrix} \right)  \;.
\end{eqnarray}
where $ \kappa = \pm 1$ and $  M_\nu $ is a $\mu-\tau$ reflection symmetric matrix. Notice that both 
$ \kappa = \pm 1 $ lead to $ \theta_{23}  = \pi/4$ \cite{Xing:2015fdg,Kitabayashi:2016zec}. This can be obtained by rearranging the phases or signs and ensuring that the mixing angles to lie in the first quadrant of the elements of V (see Eq.~\ref{eq:pmns}). In what follows we are only interested for $ \kappa = - 1 $, as our framework gives rise to $ m_{e \mu} =  - m_{e \tau}^{*}$, which is apparent from Eq.~\ref{eq:tex_1}. 

Note that in general the leptonic mixing matrix (or PMNS matrix) V ($ \equiv   U_{PMNS}$) can be expressed as $V = V^{\dagger}_lV_{\nu}$ where $ V_l, V_{\nu} $ are the unitary matrices which diagonalizes neutrino mass matrix $M_\nu^{}$ and the charged-lepton mass matrix $ M_l $, respectively. However, without loss of generality, we  choose to work in the basis where $ M_l $ is diagonal. In this scenario, $ V_l $ would reduces to a unit matrix and the leptonic mixing matrix completely determines by neutrino sector with $ V = V_{\nu}$.
In the standard PDG \cite{Patrignani:2016xqp} parameterization, one can reconstruct the elements of $  M_\nu $ using the unitary mixing matrix which  can be parameterized as
\begin{align}\label{eq:pmns}
\footnotesize
V & = P_l U P_{\nu}\; , \nonumber \\ 
 & = \left( \begin{matrix}  e^{i \phi_{e}} & 0& 0 \cr
0 & e^{i \phi_{\mu}} & 0 \cr
0 & 0 & e^{i \phi_{\tau}} \cr
\end{matrix} \right)
\left( \begin{matrix}
c^{}_{12} c^{}_{13} & s^{}_{12} c^{}_{13} & s^{}_{13} e^{-{\rm i} \delta} \cr 
 -s^{}_{12} c^{}_{23} - c^{}_{12} s^{}_{13} s^{}_{23} e^{{\rm i} \delta} & c^{}_{12} c^{}_{23} -
s^{}_{12} s^{}_{13} s^{}_{23} e^{{\rm i} \delta} & c^{}_{13}
s^{}_{23} \cr 
s^{}_{12} s^{}_{23} - c^{}_{12} s^{}_{13} c^{}_{23}
e^{{\rm i} \delta} & -c^{}_{12} s^{}_{23} - s^{}_{12} s^{}_{13}
c^{}_{23} e^{{\rm i} \delta} &  c^{}_{13} c^{}_{23} \cr
\end{matrix} \right) 
\left( \begin{matrix}  1 & 0& 0 \cr
0 & e^{i \rho} & 0 \cr
0 & 0 & e^{i \sigma} \cr
\end{matrix} \right), \;
\end{align}
where $c^{}_{ij} (s^{}_{ij}) = \cos\theta^{}_{ij} (\sin\theta^{}_{ij})$ (for $i<j=1,2,3$), 
%
$ \phi_{e, \mu, \tau}$ are the unphysical phases which can be absorbed by the rephasing of charged lepton fields ($ l_{\alpha L} , \alpha = e, \mu, \tau $)  whereas $ \rho, \sigma $ are the two Majorana phases. 
In the framework of $ \mu-\tau $ reflection symmetry, one can find six predictions~\footnote{A detailed study about the predictions of  $ \mu-\tau $ reflection symmetry has been presented in Ref.~\cite{Grimus:2003yn,Nath:2018hjx}.}, namely,
\begin{align}\label{eq:prediction}
{\rm Unphysical ~ phases} \rightarrow ~ & \phi_{e} =  0,~~~ \phi_{\mu} = - \phi_{\tau} \equiv \phi \; \nonumber \\
{\rm CP ~ phases}\rightarrow ~ & \delta =   \pm \pi/2,~~~ \rho,~\sigma = 0 ~~{\rm or}~~ \pi/2 \;, \nonumber \\
{\rm (2-3) ~ mixing ~ angle}\rightarrow ~ &  \theta_{23} = \pi/4 \;.
\end{align}
Thus, $M_{\nu}$ can be written as, 
\begin{align}\label{eq:refle_mat_1}
M_{\nu} & = V {\rm diag}(m_1, m_2, m_3)V^{T} \;,
\end{align}
where $ m_i $'s ($i = 1, 2, 3$) are the masses of light neutrinos. Also note that in the minimal seesaw formalism ${\rm det} (M_{\nu}) = 0$ which implies ${\rm det} (m_1 m_2 m_3) = 0$ and thus there is a freedom of choice to assign overall effective Majorana phase. In this study, we assign Majorana phase  $ \sigma $ with mass $m_2$. After solving eq.~(\ref{eq:refle_mat_1}), the analytical forms of low energy neutrinos mass matrix elements  under $ \mu-\tau $ reflection symmetry can be written as
\begin{eqnarray} \label{eq:LowEnergyElements}
m_{ee} & = & m_1 c_{12}^{2}c_{13}^{2} + \overline{m}_2 s_{12}^{2}c_{13}^{2} -  m_3 s_{13}^{2}\;, \nonumber \\
m_{e \mu} & = & \dfrac{e^{i \phi}}{\sqrt{2}} \left\lbrace - m_1 c_{12}c_{13}(s_{12} + i c_{12}\tilde{s}_{13})+  \overline{m}_2  s_{12}c_{13} (c_{12} - i s_{12}\tilde{s}_{13}) - i m_3 c_{13}\tilde{s}_{13}  \right\rbrace \;, \nonumber \\
m_{\mu \mu} & = & \dfrac{e^{2 i \phi}}{2} \left\lbrace  m_1 (s_{12} + i c_{12}\tilde{s}_{13})^{2}+  \overline{m}_2  (c_{12} - i s_{12}\tilde{s}_{13})^{2} +  m_3 c_{13}^{2}  \right\rbrace \;, \nonumber \\
m_{\mu \tau} & = & \dfrac{1}{2} \left\lbrace - m_1 (s_{12}^{2} + c_{12}^{2}s_{13}^{2}) - \overline{m}_2  (c_{12}^{2} +  s_{12}^{2}s_{13}^{2}) +  m_3 c_{13}^{2}  \right\rbrace \;, 
\end{eqnarray}
where $ \overline{m}_2 = \pm m_2 $ and `$ \pm $' is for $  \sigma = 0~ {\rm or} ~\pi/2 $, also $ \tilde{s}_{13} = \pm s_{13} $ for $  \delta = \pm\pi/2 $. Note that in the minimal seesaw formalism  $  m_1 (m_3) = 0 $ for NO (IO).

The low energy neutrino mass matrix $ M_{\nu} $ as given by eq.(\ref{eq:low_mnu}) and its elements as mentioned in eq.(\ref{eq:LowEnergyElements}) can also be considered as a consequence of the type - I seesaw mechanism. In this context, we consider different scenarios of Dirac neutrino mass matrix,  $ M_D $ as conjectured in littlest seesaw model as \cite{King:2015aea,King:2016yef}, 
\begin{eqnarray}\label{eq:md_LSS}
{\rm SI : ~~} M_D &=  \left( \begin{matrix} 0&  b e^{i \eta/2} \cr
 a  &  n b e^{i \eta/2} \cr  
 a &   (n-2) b e^{i \eta/2}
\end{matrix} \right), ~~~
{\rm SII : ~~} M_D =  \left( \begin{matrix} 0&  b e^{i \eta/2} \cr
 a  &  (n-2) b e^{i \eta/2} \cr  
 a &   n b e^{i \eta/2}
\end{matrix} \right) \nonumber \;, \nonumber \\
{\rm SIII : ~~} M_D &=  \left( \begin{matrix}   b e^{i \eta/2} & 0\cr
  n b e^{i \eta/2} & a\cr  
   (n-2) b e^{i \eta/2} & a
\end{matrix} \right), ~~~
{\rm SIV : ~~} M_D =  \left( \begin{matrix}   b e^{i \eta/2} & 0 \cr
  (n-2) b e^{i \eta/2} & a\cr  
  n b e^{i \eta/2} & a
\end{matrix} \right)\;,
\end{eqnarray}
whereas we consider Majorana neutrino mass matrix $ M_R $  as
\begin{eqnarray}\label{eq:mr}
M_R = \left( \begin{matrix} r_1 e^{i \phi_{m}}  & r_2\cr
   r_2 & r_1 e^{-i \phi_{m}} 
\end{matrix} \right) \;.  \end{eqnarray}
Now, considering $n = 1 ~~ {\rm and} ~~ \eta = (2 l + 1) \pi ~ (l = 0, 1, 2,...)$, we obtain 
$M_D$ for SI as
\begin{equation}\label{eq:md}
 M_D =  \left( \begin{matrix} 0&  i b  \cr
 a  &  i b  \cr  
 a &   - i b 
\end{matrix} \right) \;.
\end{equation}
For $ \phi_{m} = (2 m + 1)\pi ~{\rm with} ~ m = 0, 1, 2, ...   $, Majorana neutrino mass matrix becomes \footnote{Note that $ \phi_{m} = 2 m \pi ~{\rm with} ~ m = 0, 1, 2, ...   $, changes $ r_1 \rightarrow - r_1$, leading to an overall sign difference in eq.(\ref{eq:tex_1}).},
\begin{eqnarray}\label{eq:tex_1}
M_R & = & \left( \begin{matrix} - r_1   & r_2\cr
   r_2 & - r_1 
\end{matrix} \right) \;. \end{eqnarray}
We like to emphasize here that, in principle, one can construct these mass textures within some suitable flavor symmetry formalism along with preferred $ \mathcal{Z}_n $ group. In what follows, without loss of generality, we proceed to study the phenomenology of these textures rather their theoretical origin
\footnote{Recently,  the $ \mu-\tau $ reflection symmetric $ M_{\nu} $ within the littlest seesaw formalism under $ S_4 $ flavor group has been discussed in Ref.\cite{King:2019tbt}.}.

Given the above  $( 2\times2 )$ matrix $ M_R $ (eq.\ref{eq:tex_1}), one can always diagonalize it using a trivial mixing matrix of the form,
\begin{eqnarray}\label{eq:u23}
U_{23}& = &  \dfrac{1}{\sqrt{2}}\left( \begin{matrix} 1   & 1\cr
   -1 & 1 
\end{matrix} \right) \;. \end{eqnarray}
Thus $ M_R $ becomes,
\begin{eqnarray}\label{eq:MrHat}
\widehat{M}_R& = &  U^{T}_{23} M_R U_{23} \;, \nonumber  \\
 & = &\left( \begin{matrix} - r_1 - r_2   & 0\cr
   0 & r_2 - r_1 
\end{matrix} \right) \;. \end{eqnarray}
This transformation changes $ M_D $ to be
\begin{equation}\label{eq:MdHat}
 \widehat{M}_D = \dfrac{1}{\sqrt{2}}  \left( \begin{matrix}
 - i b &  i b  \cr
 a - i b &  a+ i b  \cr  
 a + i b&   a - i b 
\end{matrix} \right) \;.
\end{equation}
Notice here that in principle one can start with these particular textures of $ \widehat{M}_R, \widehat{M}_D$ without concerning about their theoretical origin.
Moroever,  one can construct the low energy neutrino mass matrix $ M_{\nu} $ under the type-I seesaw mechanism  with these textures of $ \widehat{M}_R, \widehat{M}_D$  that satisfies $ \mu - \tau $ reflection symmetry as 
%
%
\begin{eqnarray}\label{eq:typeIseesaw}
 - M_{\nu} & = & \widehat{M}_D  \widehat{M}_R^{-1}  \widehat{M}_D^T \;, \\
 & = &  \dfrac{1}{{r_1^2-r_2^2}}  \left( \begin{array}{ccc}
 b^2 r_1 & (b^2 r_1 - i a b r_2) & -(b^2 r_1 + i a b r_2) \\
* & -\left(a^2-b^2\right) r_1-2 i a b r_2 & -\left(a^2+b^2\right)r_1 \\
* & * & -\left(a^2-b^2\right) r_1 + 2 i ab r_2 \\
\end{array}
\right) \;,
\end{eqnarray}
%
%
%
The noteworthy output of this texture is that it predicts maximal atmospheric mixing angle along with maximal value of leptonic CP violating phase which are in good agreement with the latest oscillation results \cite{Capozzi:2016rtj,Esteban:2016qun,deSalas:2017kay}.
Notic here the importance of off-diagonal term of eq.(\ref{eq:mr}) as $ r_2 \rightarrow 0 $, $ M_{\nu}  $ becomes a real symmetric matrix which obeys $ \mu - \tau $ permutation symmetry \cite{Fukuyama:1997ky} and leads to a vanishing mixing angle $ \theta_{13} $ which is ruled out by current data.
One can also write eq.(\ref{eq:tex_1}) as 
\begin{eqnarray}\label{eq:MnuAfterSeesaw}
 M_{\nu} & = &  \left( \begin{matrix}A & B &  - B^{*} \cr
  * & C & D \cr
 * & * & C^{\ast} \cr
\end{matrix} \right), 
\end{eqnarray}
where $ A, D $ are real and 
\begin{eqnarray}\label{eq:low_high}
A & = &  b^2 r^{\prime}_1 \;,  \nonumber \\
B & = & (b^2 r^{\prime}_1 - i a b r^{\prime}_2)\;,  \nonumber \\
C & = & -\left(a^2-b^2\right) r^{\prime}_1 - 2 i a b r^{\prime}_2\;,  \nonumber \\
D & = & -\left(a^2+b^2\right)r^{\prime}_1 \;, 
\end{eqnarray}
with $ r^{\prime}_{1,2} = r_{1,2}/(r_1^2-r_2^2) $ \footnote{The analogous form of $ M_{\nu} $ as given by eq.(\ref{eq:tex_1}) can also be constructed by considering other scenarios (SII, SIII, SIV) of $ M_D $ which we will discuss in Appendix~\ref{sec:appendix}.}. At low energy, one can make one to one correspondence among the matrix elements of eq.(\ref{eq:MnuAfterSeesaw}) and eq.(\ref{eq:low_mnu}).
Further, we find that the matrix  elements of $  M_{\nu} $ are not fully independent and they are correlated as
\begin{equation}\label{eq:cond}
{\rm Re} (B) = A\;, ~2{\rm Im} (B)   =  {\rm Im} (C) \;, ~ 2{\rm Re} ( B )+ D = {\rm Re}  (C) \;.
\end{equation}
Imposing the conditions mentioned in eq.(\ref{eq:cond}), one can re-write high energy neutrino mass matrix elements in terms of low energy parameters of eq.(\ref{eq:low_high}) as
\begin{eqnarray}\label{eq:CorrelationHiLow}
\dfrac{b^2}{a^2} & = & \left|  \dfrac{A}{(A+D)} \right| \;,  \nonumber \\
 \dfrac{r^{\prime}_1}{r^{\prime}_2} & = &   \dfrac{r_1}{r_2} = \left| \dfrac{2}{{\rm Im}(C)}\sqrt{-A(A+D)} \right| \;.
\end{eqnarray}

In order to better understand of high energy parameters in terms of low energy neutrino mass matrix elements, we make expansion of eq.(\ref{eq:CorrelationHiLow}) in the framework of $ \mu-\tau  $ reflection  symmetry. Considering two small parameter, $ \zeta = m_2/m_3 $ and $ \theta_{13} $, we find
\begin{align}\label{eq:CorrelationHiLowFinal}
\dfrac{b^2}{a^2}  \simeq ~ & -  2 \zeta s_{12}^{2} + 2 \theta_{13}^{2} - \zeta  \theta_{13}^{2}(3 - 5 \cos 2 \theta_{12}) + \mathcal{O}(\zeta^{2}\theta_{13}^{2}) \;,\nonumber \\
 \dfrac{r_1}{r_2} \simeq ~ &  \dfrac{2 s_{12} \sqrt{2 \zeta} }{\sin 2 \phi} -\frac{\zeta^{3/2} (1 + 5 \cos2 \theta_{12}) s_{12}}{\sqrt{2} \sin 2 \phi }  - \frac{\sqrt{2} \theta_{13}^{2}}{\sqrt{\zeta} s_{12} \sin 2 \phi}  \nonumber \\
  & - \dfrac{4\sqrt{2} \theta_{13} \zeta^{3/2}c_{12}s^{2}_{12}\cos 2 \phi }{\sin^{2} 2 \phi} + \mathcal{O}(\zeta^{2}\theta_{13}^{2})\;.
\end{align}

On the other hand, as we notice that $\mu-\tau  $ reflection symmetry does not shed any light on the mixing angles, $   \theta^{}_{12}, \theta^{}_{13} $, we express the mixing angles in terms of model parameters as
\begin{align}
 \theta^{}_{13} & = \mp ~  
\arctan \left[ \frac{1}{\sqrt{2}} \frac{{\rm Im} \left(C^{\prime}\right)}
{{\rm Re} \left(B^{\prime}\right) } \right]  ~;  \nonumber \\
& = \mp ~  
\arctan \left[  \frac{1}{\sqrt{2}} \frac{(a^2 - b^2)r^{\prime}_1 \sin2\phi - 2abr^{\prime}_2 \cos2\phi }{b^2 r_1^{\prime}\cos \phi - abr^{\prime}_2 \sin \phi  }  \right]~;  \nonumber \\
\theta^{}_{12} &= 
\dfrac{1}{2} \arctan \left[-   \dfrac{2\sqrt{2} \cos 2\theta_{13} {\rm Re} \left(B^{\prime}\right)}{c_{13}\left[{\rm Re(C^{\prime})}(s^{2}_{13} - \cos 2\theta_{13}) +  D (s^{2}_{13} + \cos 2\theta_{13})  + A  c^{2}_{13}\right] } \right]\;\nonumber \\
 &= 
\dfrac{1}{2} \arctan \left[ -  \frac{2\sqrt{2} \cos 2\theta_{13} (b^2 r_1^{\prime}\cos \phi - abr^{\prime}_2 \sin \phi) }
{\splitfrac{ c_{13}[ (b^2 - a^2)r^{\prime}_1 \cos2\phi - 2abr^{\prime}_2 \sin2\phi)(s^{2}_{13} - \cos 2\theta_{13})}
{-(a^2 + b^2)r^{\prime}_1 (s^{2}_{13} + \cos 2\theta_{13})  + b^2 r^{\prime}_1  c^{2}_{13} ] }} \right] \;,
\end{align}
where,  $C^{\prime}=C e^{-2i\phi}, B^{\prime}=B e^{-i \phi} $ and $` \mp $' represents the sign corresponding to the value of $ \delta = \pm \pi/2$. Moreover, in this scenario we calculate $ \phi $ in terms of model parameter as,
\begin{equation}
\phi = \dfrac{1}{2} \arctan[\pm (b^2 - a^2) r_1 , \mp 2 a b r_2  ] + 2 \pi c_1, ~~ c_1 \mathbb{\epsilon} ~ \mathbb{Z}.
\end{equation}
Furthermore, the masses of the light neutrinos corresponding to NO can be calculated as,
\begin{align}
m_1 = & ~0 \;, \nonumber \\
m_2 e^{2 i \sigma}=  &~   \dfrac{2\sqrt{2} {\rm Re}(B^{\prime}) }{c_{13}\sin 2 \theta_{12}} \;  , \nonumber \\
= & ~  \dfrac{2\sqrt{2}(b^2 r_1^{\prime}\cos \phi - abr^{\prime}_2 \sin \phi) }{c_{13}\sin 2 \theta_{12}} \;, \nonumber \\
m_3 = &~  \dfrac{2\sqrt{2} {\rm Re}(B^{\prime}) }{c_{13}\sin 2 \theta_{12}}+ 2D - A \, , \nonumber \\
= & ~ \dfrac{2\sqrt{2}(b^2 r_1^{\prime}\cos \phi - abr^{\prime}_2 \sin \phi) }{c_{13}\sin 2 \theta_{12}} - 2 (a^2 + b^2)r^{\prime}_1 - b^2 r^{\prime}_1 \;.
\end{align}

Before proceeding further with the phenomenological descriptions, we first present here the simulation details that have been carried out throughout this paper.
We also like to emphasize  that  while performing numerical analysis, we do not assume any  approximations. All the analyses are based on exact formulation. We treat  both the parameters ($ a, b $) of $M_D$ and  ($r_1, r_2  $) of $M_R^{}$ as free parameters in the numerical simulation and vary them as
\begin{eqnarray}\label{eq:VaryParam}
a, b \in [-1, 1]~v, ~~  \quad r_1, r_2 \in [10^{12}, 10^{15}]~\mathrm{GeV} \;.
\end{eqnarray}
While performing numerical analysis, we use the nested sampling package $\texttt{Multinest}$ \cite{Feroz:2007kg,Feroz:2008xx,Feroz:2013hea} for the parameter scan with an assigned $\chi^2$ function  based on the latest global-fit analysis of neutrino oscillation data~\cite{deSalas:2017kay}. 
The Gaussian-$\chi^2$ function that we use in our numerical simulation is defined as,
\begin{equation}
\chi^{2} = \sum_i \dfrac{\left[  \xi_i^{\rm True} - \xi_i^{\rm Test} \right] ^{2}  }{\sigma \left[ \xi_i^{\rm True} \right] ^{2}} \;,
\end{equation}
where  $\xi = \{ \theta_{ij}, \Delta m_{21}^{2}, |\Delta m_{31}^{2}| \}$ with ($ ij = 12, 13, 23 $),  represents the set of neutrino oscillation parameters. Also, $\xi_i^{\mathrm{Ture}}$ represent the current best-fit values from the global analysis of neutrino oscillation data~\cite{deSalas:2017kay} whereas $\xi_i^{\mathrm{Test}}$ correspond to the predicted values for a given set of parameters in theory. We also  symmetrize standard deviation, $\sigma \left[ \xi_i^{\rm True} \right]$ considering 1$\sigma$ errors as given by Ref.~\cite{deSalas:2017kay}. In this study, we collect all the scattered points which have $ \chi^{2} < 30 $.
Further, in the numerical analysis, we notice that for the inverted neutrino mass ordering scenario, this model gives few scatter points with $  \chi^{2}_{min} > 100 $ with no definite correlations.  Thus, we conclude that the given model disfavors inverted neutrino mass ordering. This is also 
in well agreement with the prediction of littlest seesaw model \cite{Antusch:2011ic,King:2013xba,King:2013iva,Bjorkeroth:2014vha,King:2015aea,King:2015dvf,Bjorkeroth:2015tsa,King:2016yvg,King:2016yef,King:2018kka,King:2018fqh,Ding:2018fyz}.

\begin{figure}[h!]
\centering
\includegraphics[height=7cm,width=15cm]{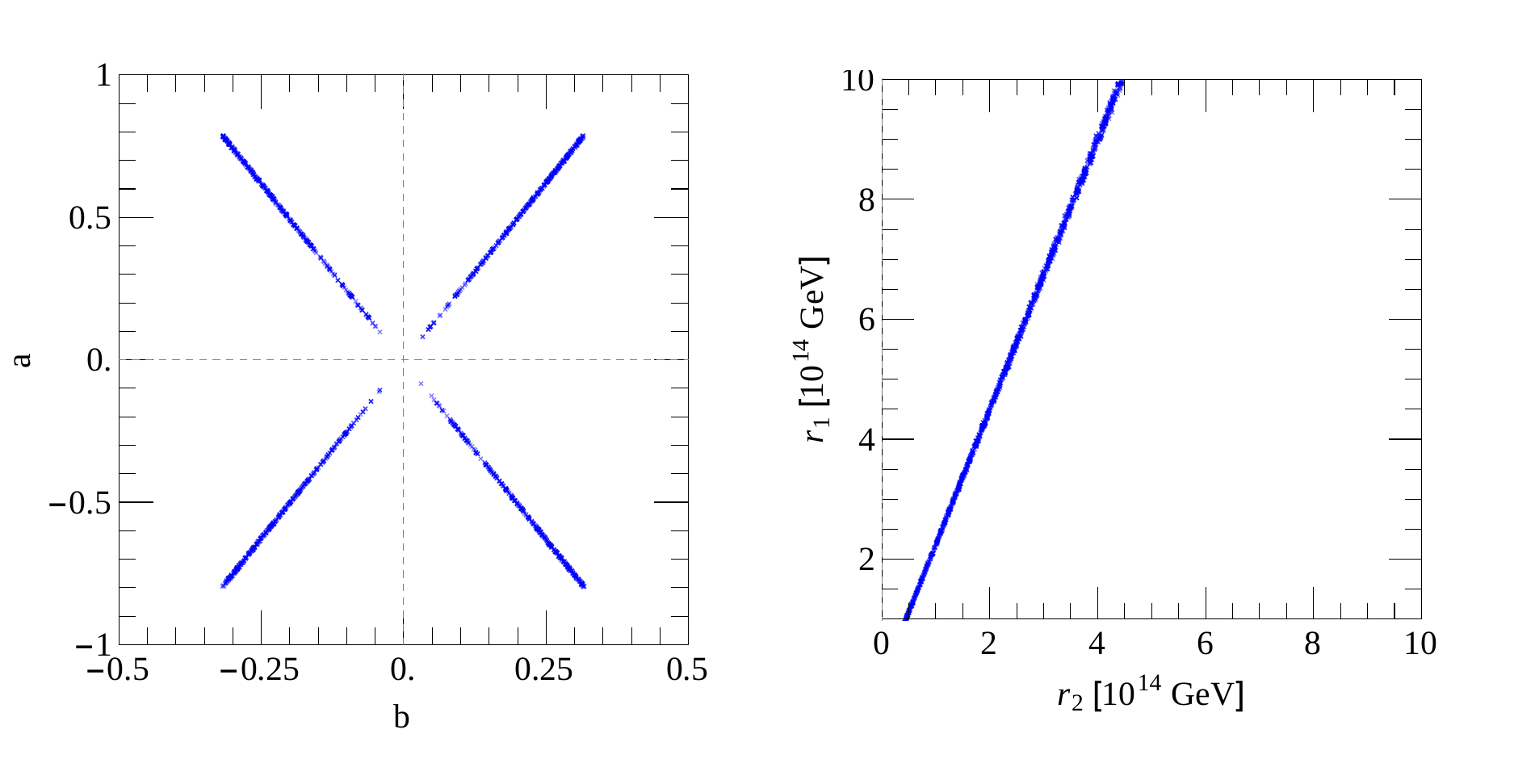}
\caption{\footnotesize Correlation plots among high energy neutrino mass matrix elements.}
\label{fig:corrNH}
\end{figure}

In fig.(\ref{fig:corrNH}), we describe various correlations among high energy neutrino mass matrix elements within the framework of $\mu-\tau  $ reflection symmetry.
Using the current best-fit values of neutrino oscillation data \cite{deSalas:2017kay}, we find $ b/a\simeq 0.26, r_1/r_2 \simeq 2.1 $ for $ \zeta \simeq 0.17 $, $ \theta_{13} \simeq 0.15$  and $ \phi \simeq 185^\circ $ from eqs.~(\ref{eq:CorrelationHiLowFinal}). These results are in well agreement with our numerical results.
Further, we observe a very strong correlation between the elements of $ M_D $ i.e., among $ a$  and $b$ as shown by left panel. From this figure, we notice an important point that  neither of these elements can be zero. This is simply because if any of these elements become zero then it will 
 lead to two massless neutrinos which is ruled out by current neutrino oscillation data.
Moreover, a positive correlation between the elements of $ M_R $ has been observed from  right panel.
%

Having discussed the novel features of $ \mu - \tau $ reflection under the minimal seesaw model, we now proceed to address another intriguing aspect of particle physics and cosmology, namely the Baryon asymmetry of the Universe (BAU). In next section, a detailed study on BAU arising from the asymmetry between leptons and antileptons, vastly known as \textit{leptogenesis}, has been performed.
\section{Leptogenesis}\label{sec:Leptogenesis}
The lepton asymmetry  serves as a preeminent mechanism to understand the excessive nature of baryonic matter over antimatter in the observable universe. Further, this lepton asymmetry can be converted into the Baryon asymmetry of the Universe through sphaleron process \cite{Buchmuller:2005eh}.
In order to  have observable baryon asymmetry, Sakharov proposed three principle ingredients  that a baryon-generating interaction of the universe must satisfy to generate matter and antimatter at different rates \cite{Sakharov:1967dj}. These conditions are called as Sakharov conditions namely, (i) baryon (B) number violation, (ii) charge conjugation (C) and charge-parity (CP) violation, and (iii) out of equilibrium dynamics. Moreover, one can express  the baryon
asymmetry parameter, $ Y_B $, as the ratio of difference in number densities of baryons ($ n_B $) and antibaryons ($ n_{\overline{B}} $) to the entropy density (${\rm s}$) of the universe \cite{Buchmuller:2005eh} as
\begin{equation}
Y_B \equiv \dfrac{n_B - n_{\overline{B}}}{{\rm s}} \;,
\end{equation}
which is also known as the final baryon-to-entropy ratio.
The current experimentally observed value of  $ Y_B $ lies in the range, $ 8.55 \times 10^{-11} < Y_B < 8.77\times10^{-11} $ \cite{Ade:2015xua}.

It is well known that the CP violation at the low energy can be correlated with the CP violation 
at the high energy in the seesaw framework~\cite{Endoh:2002wm,Branco:2002xf,Pascoli:2006ci}. The CP violating, out of equilibrium and  lepton-number-violating  decays of heavy right-handed Majorana neutrinos ($ N_i $) provide a natural way to explain the lepton asymmetry of the universe which was first proposed in Ref.~\cite{Fukugita:1986hr} (see Ref.~\cite{Davidson:2008bu} for a  review and the references therein).  The asymmetry generated due to the decay of lightest heavy neutrino $N_1$  into the lepton doublet $ l_{\alpha }$ corresponding to flavor $ \alpha (\equiv e, \mu, \tau)$ and Higgs doublet $ H $ can be expressed as \cite{Xing:2011zza}
\begin{equation}\label{eq:CPAsymmetry}
\pmb{\bm{\varepsilon}}_{\alpha}^{1} \equiv  \dfrac{\Gamma(N_1 \rightarrow H l_{\alpha}) - \Gamma(  N_1 \rightarrow H^{\dagger} \overline{l}_{\alpha}) }{\Gamma(N_1 \rightarrow H l_{\alpha}) + \Gamma(N_1 \rightarrow H^{\dagger} \overline{l}_{\alpha})} \;,
\end{equation}
where $ \overline{l}_{\alpha} $ are the leptonic fields corresponding to antiparticles.
  One can calculate this asymmetry in terms of Dirac neutrino Yukawa couplings in diagonal Majorana neutrino mass basis as \cite{Covi:1996wh,Plumacher:1996kc,Xing:2011zza}
\begin{equation}
\pmb{\bm{\varepsilon}}_{\alpha} \simeq  \dfrac{1}{8 \pi}  \dfrac{{\rm Im} [ (Y_D^{\dagger} Y_D)_{12} (Y_D^{\dagger} )_{1 \alpha} (Y_D)_{\alpha 2} ] }{(Y_D^{\dagger} Y_D)_{11} } f\left( \dfrac{M^2_2}{M^2_1}\right) \;.
\end{equation}
Note that now onwards, we neglect super-script `1' from $ \pmb{\bm{\varepsilon}}^{1}_{\alpha} $ for simplicity. Here $f(x)$ is the loop function which includes one-loop vertex and self-energy corrections of the decay amplitude of right-handed neutrino field and is given by \cite{Xing:2011zza},
\begin{equation}
f(x) = \sqrt{x} \left[(1+x) {\rm ln}\left(\dfrac{x}{1+x} \right) + \dfrac{2 - x}{1 + x} \right] \;.
\end{equation}
In the hierarchal limit of heavy right-handed Majorana neutrino masses, $ M_1 \ll M_2$ (i.e., $ x =  \dfrac{M^{2}_2}{M^{2}_1}\gg 1 $), one can approximate $ f(x)  $ \cite{Davidson:2008bu} as
\begin{equation}
f(x) \simeq - \dfrac{3}{2\sqrt{x}} - \dfrac{5}{6 x^{3/2}} + ...  \;
\end{equation}
  Using the first order approximation, we find the CP asymmetry in our case as \cite{Xing:2011zza}
\footnote{We like to mention here that during the leptogenesis epoch, the electroweak symmetry has not been broken and the concerned symmetry act on lepton doublets. To build a consistent model with $\mu-\tau$ flavor symmetry, authors of Ref.(\cite{Mohapatra:2015gwa}) have pointed out that one can construct a model starting with a high scale symmetry where low energy effective theory i.e., neutrino sector still maintains  $\mu-\tau$ flavor symmetry, on the other hand symmetry is spontaneously broken in the charged-lepton sector to allow non-degenerate masses for the muon and tau leptons. The main challenge to build such model is to keep unbroken  $\mu-\tau$ flavor symmetry for the neutrino sector whereas break it for the charged-lepton sector. Authors have added two more Higgs doublets in the model (named as gauged $ U(1)_{\mu-\tau} $) which couples with muon and tau sector differently. Further, successful leptogenesis has also  been discussed in that scenario.}
\begin{equation}\label{eq:FinalCPAsymmetry}
\pmb{\bm{\varepsilon}}_{\alpha} \simeq - \dfrac{3}{16 \pi}  \dfrac{{\rm Im} [ (Y_D^{\dagger} Y_D)_{12} (Y_D^{\dagger} )_{1 \alpha} (Y_D)_{\alpha 2} ] }{(Y_D^{\dagger} Y_D)_{11} } \dfrac{M_1}{M_2}\;.
\end{equation}

Now, we calculate the CP asymmetry, $\pmb{\bm{\varepsilon}}_{\alpha}$ in this model using $ M_D (\equiv v Y_D) $ as given by eq.(\ref{eq:md}). We find vanishing resultant CP asymmetry i.e., $ \pmb{\bm{\varepsilon}} = \sum \pmb{\bm{\varepsilon}}_{\alpha} = 0$ with $ \pmb{\bm{\varepsilon}}_e  = 0$ and $ \pmb{\bm{\varepsilon}}_{\mu}  = - \pmb{\bm{\varepsilon}}_{\tau}$. Nonetheless, one can find non-zero CP asymmetry in the broken scenario of $ \mu - \tau $ reflection symmetry~\cite{Ahn:2008hy}.
 Introducing explicit breaking term, $ \epsilon $, in the (3,1) position of $ M_D $ as given by eq.(\ref{eq:md}), one can write the modified form of $ M_D $ as,
\begin{equation}\label{eq:MdPrime}
M_D^{\prime} =  \left(
\begin{array}{cc}
0 & i b \\
a  & i b \\
 a (1 + \epsilon)  &  - i b \\
\end{array}
\right) \;,
\end{equation}
which in the diagonal Majorana neutrino mass basis becomes 
\begin{equation}
\widehat{M}_D^{\prime} = \frac{1}{\sqrt{2}} \left(
\begin{array}{cc}
- i b & i b \\
a - i b & a + i b \\
 a (1 + \epsilon) + i b &  a (1 + \epsilon) - i b \\
\end{array}
\right) \;,
\end{equation}
where $ \epsilon $ is the symmetry breaking parameter.
Notice that non-zero $ \epsilon $ leads to $ (\widehat{M}_D^{\prime})_{\mu i} \neq (\widehat{M}_D^{\prime *})_{\tau i} $ (where i = 1, 2) that eventually reflects to $ M_{\nu} $ and break the $ \mu-\tau $ reflection symmetry, which we will discuss in section~\ref{sec:SymmetryBreaking}. Furthermore, this also leads to non-zero CP asymmetry as can be seen from eq.(\ref{eq:CPAssy}).
Given the form of $ \widehat{M}_D^{\prime} $, one finds
\begin{equation}
\widehat{M}_D^{\prime \dagger} \widehat{M}_D^{\prime} =  \frac{1}{2 } \left(
\begin{array}{cc}
2 (1 +\epsilon) a^2+3 b^2 & 2 (1 + \epsilon) a^2-2 i a b \epsilon  -3b^2 \\
2 (1 +\epsilon) a^2+2 i a b \epsilon  -3 b^2 & 2 (1 +\epsilon) a^2+3b^2 \\
\end{array}
\right) \;.
\end{equation}

Employing the above form, we calculate the final CP asymmetry
 by following eq.(\ref{eq:FinalCPAsymmetry}) as,
\begin{align}\label{eq:CPAssy}
\pmb{\bm{\varepsilon}} = & ~ -\dfrac{3}{16 \pi} \dfrac{M_1}{M_2} \sum_\alpha \pmb{\bm{\varepsilon}}_{\alpha} \;, \nonumber \\
 = & ~ \epsilon ~ \dfrac{3}{16 \pi} \dfrac{M_1}{M_2} \frac{2 a b    \left(2 a^2 - 3 b^2\right)}{v^2(2 a^2+3 b^2)} \;,
\end{align}
where we used $Y_D^{\prime} = M_D^{\prime}/v $. This asymmetry can be related to the final baryon asymmetry as \cite{Buchmuller:2005eh}
 \begin{equation} 
  Y_B \equiv  \dfrac{n_B - n_{\overline{B}}}{{\rm s} } \simeq - \dfrac{12}{37} \kappa \dfrac{\pmb{\bm{\varepsilon}} }{g_{*}} \;. 
 \end{equation}
Here, 12/37 is the fraction of lepton asymmetry converted into baryon asymmetry through sphaleron processes and $g_{*} = 106.75$ is the effective number of relativistic degrees of freedom which contribute to  entropy of the Universe  in the SM. Note also that $ \kappa $ is the washout factor which can be parametrized as \cite{Buchmuller:2005eh}
\begin{equation}
\kappa \simeq (2 \pm 1)\times  0.02 \times \left( \dfrac{0.01 {\rm eV}}{\tilde{m}_1}  \right) ^{1.1 \pm0.1} \;.
\end{equation}
%

Further, we calculate the effective neutrino mass parameter in our model as 
\begin{align}
\tilde{m}_1 = & ~\dfrac{(\widehat{M}_D^{\prime \dagger} \widehat{M}_D^{\prime} )_{11} }{M_1} \;, \nonumber \\ 
= & ~\dfrac{ [(1 + \epsilon ) a^2+3 b^2/2] }{M_1}\;.
\end{align}


\begin{figure}[h!]
\hspace{-0.5cm}
\includegraphics[height=7cm,width=17cm]{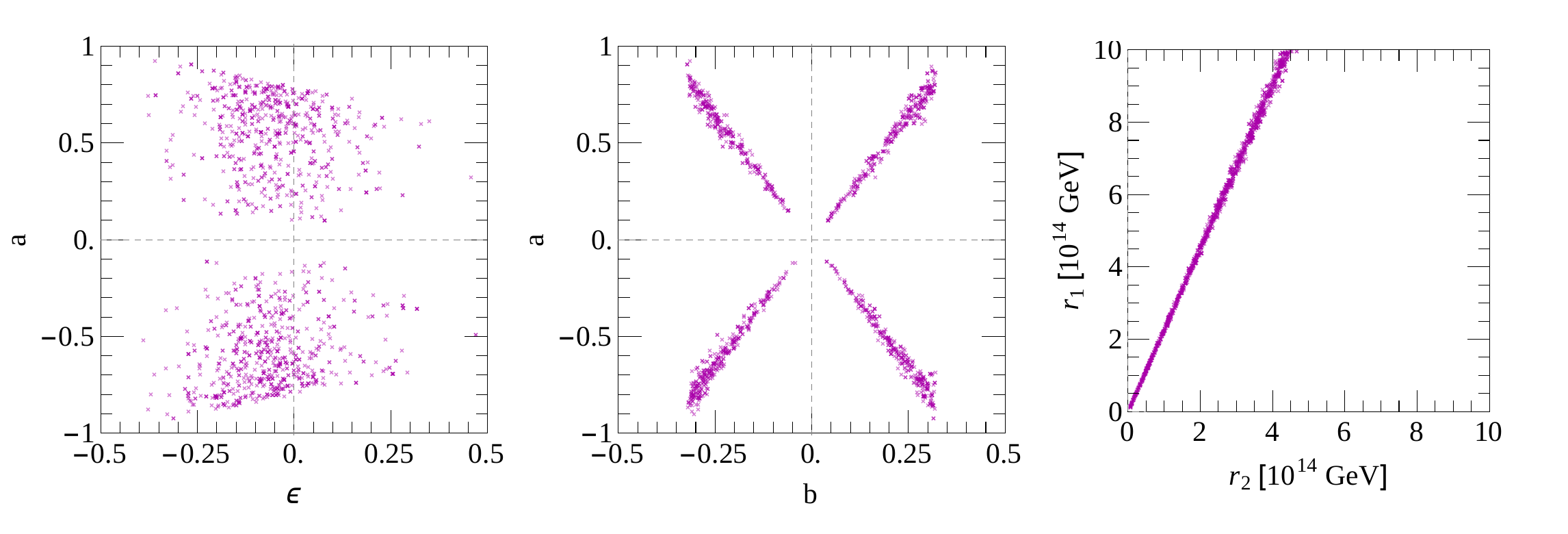}
\caption{\footnotesize Correlation between high energy neutrino mass matrix elements. Here, all the magenta points have $ \chi^{2} < 30 $.}
\label{fig:NuOscParam_NH_HE}
\end{figure}
Having discussed the theory behind the baryogenesis via leptogenesis, we now proceed to analyze the phenomenology of lepton-antilepton asymmetry in the given model. In fig.~(\ref{fig:NuOscParam_NH_HE}), we demonstrate different correlations among high energy parameters which can explain current neutrino oscillation data in their 3$ \sigma $ ranges.
First panel shows the correlation between model parameter, $a$, with the symmetry breaking parameter,  $ \epsilon $. We notice from this figure that a small value of non-zero $ \epsilon $ can lead to non zero CP- asymmetry as given by eq.~(\ref{eq:CPAssy}). We also observe that breaking parameter, $ \epsilon $ lies in the region ( $ - 0.4 \rightarrow $ 0.35) whereas most favored parameter space is $ \sim -$ 0.15.
  From second and third plot, we notice a linear relation between the parameters of $ \widehat{M}^\prime_D $, ($ a, b $) and the parameters of $ \widehat{M}_R $, ($ r_1, r_2 $). Comparing these two plots with fig.~(\ref{fig:corrNH}), we observe mild widening among different parameters in the latter case due to the deviation from exact $ \mu - \tau $ reflection symmetry. We also find  model parameter $ |a| $ falls in the range $(0.1 - 0.8) v$ and $ |b| $ lies in $(0.05 - 0.35) v$ whereas both the variables of $ M_R $ falls around $ \sim 10^{14} $ GeV to explain latest oscillation results.
  


As has been pointed out in Refs.~\cite{Buchmuller:2005eh,Davidson:2008bu} that depending on the temperature regime in which leptogenesis takes place, one can have fully distinguishable, partly distinguishable/indistinguishable  lepton flavors.
 In order to understand temperature regime in the this model (which is equivalent to find  parameter space of the masses of heavy right-handed neutrinos), 
we find here that $ T \sim M_1 \geq 10^{13} $ GeV, all the flavors are indistinguishable and the final total CP asymmetry is given by sum of all the three flavors i.e., $ \pmb{\bm{\varepsilon}} = \sum \pmb{\bm{\varepsilon}}_{\alpha}$\cite{Buchmuller:2005eh,Davidson:2008bu}. The analytical expression corresponding to this scenario is given by eq.(\ref{eq:CPAssy}).

%


\begin{figure}[h!]
\centering
\includegraphics[height=14cm,width=13cm]{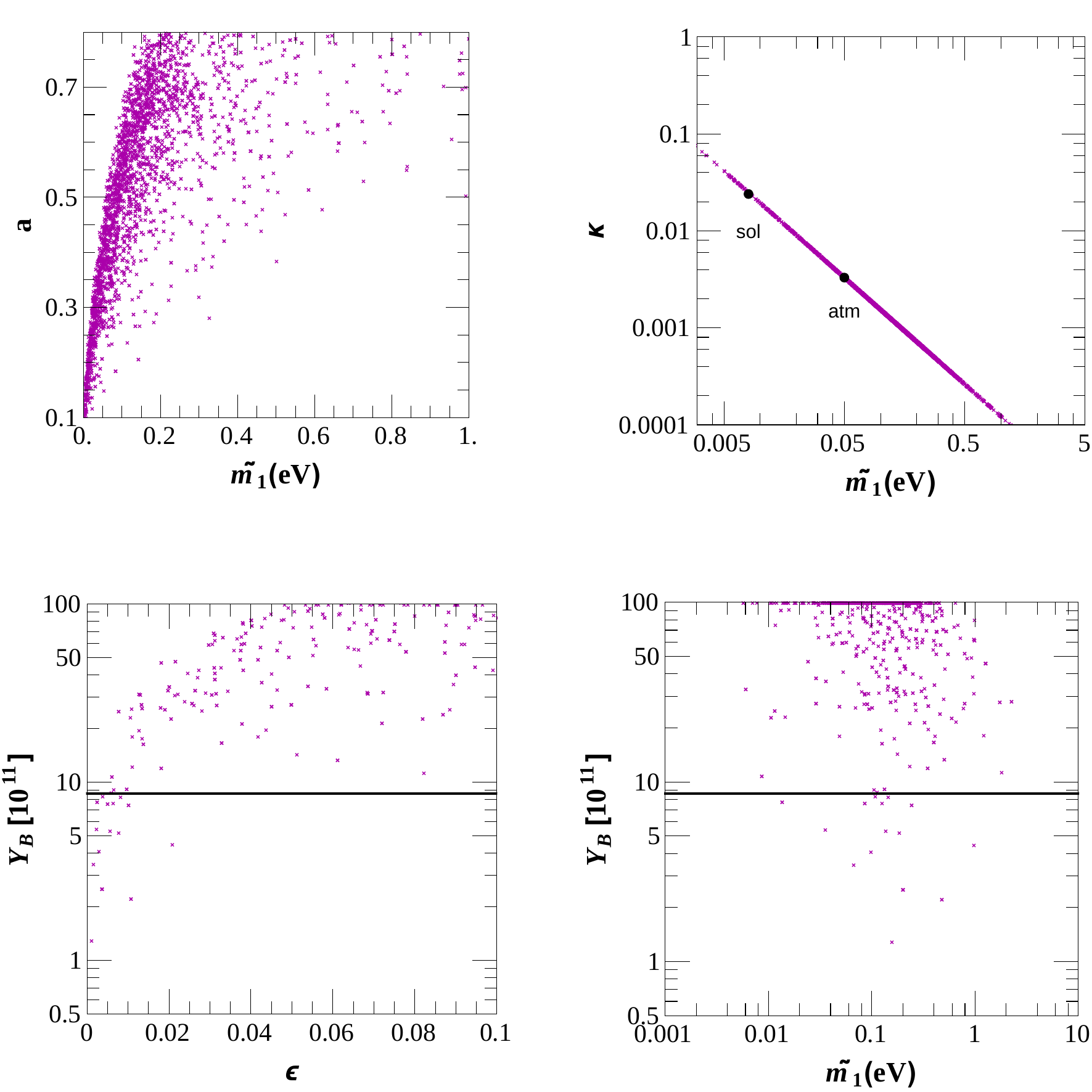}
\caption{\footnotesize Top row shows different observables which explain possible leptogenesis in our model. Whereas bottom row explains the observed BAU of our model where black tiny band corresponds to current experimental bound.}
\label{fig:Leptogenesis}
\end{figure}
In fig.(\ref{fig:Leptogenesis}), we illustrate our numerical results for leptogenesis  in substantial details. To measure 
different observables related to leptogenesis, we consider allowed parameter space of different high energy neutrino mass matrix elements (i.e., $ a, b $) which are consistent with allowed range of neutrino oscillation data as given by fig.~(\ref{fig:NuOscParam_NH_HE}) whereas input values of the lightest heavy Majorana neutrino mass is considered  $ M_1 \geq 10^{13} $ GeV. From first row, we observe that the effective neutrino mass parameter, $ \tilde{m}_1  $ lies in the region $ 10^{-2} ({\rm eV}) < \tilde{m}_1 < 1 ({\rm eV}) $ which falls in strong washout regions. In left plot, we show the behavior of model parameter, $ a $ as a function of $  \tilde{m}_1$. On the other hand, we show wash out factor ($ \kappa $) as a function of $  \tilde{m}_1$ where black dots are the current global-fit results corresponding to $  \tilde{m}_1 \equiv m_{\rm sol} \sim \sqrt{\Delta m^{2}_{21}}$ (as marked as ``sol") and $  \tilde{m}_1 \equiv m_{\rm atm} \sim \sqrt{|\Delta m^{2}_{31}|}$ (as marked as ``atm") in  right plot.

  Further, we proceed to measure the amount of baryon asymmetry ($ Y_B $) in the given framework as shown in bottom row. We show baryon asymmetry as a function of  symmetry breaking parameter $ \epsilon $ as well as with effective neutrino mass $ \tilde{m}_1  $. From, first plot of bottom row we notice that a tiny breaking term of the order of $ \sim 10^{-2} $ is able to generate sizeable amount of baryon asymmetry. We find several points which satisfy current experimental bound on baryon asymmetry (as shown by narrow black band) from both the plots of bottom row. 

After presenting a general discussion on leptogenesis under the concerned seesaw formalism, in next section we discuss the  impact of breaking term on different neutrino oscillation parameters. 

\section{Phenomenology with broken $ \mu-\tau  $ reflection symmetry}\label{sec:SymmetryBreaking}
In this section, we perform various correlation studies among three flavor neutrino oscillation parameters given the framework of broken $ \mu-\tau  $ reflection symmetry. Authors of
Refs.~\cite{Liu:2017frs,Zhao:2017yvw,Nath:2018hjx} have performed a detailed study on the breaking of such an exact symmetry considering both one-loop renormalization group (RG) running as well as by introducing an explicit breaking term in the high energy neutrino mass matrices. However,  a mild deviations from the exact symmetry has been observed considering RG running. Keeping this in mind, we establish here a general set-up to break $ \mu-\tau  $ reflection symmetry by an explicit breaking term in the Dirac neutrino mass matrix $ M_D $.  As one can notice that there exists a  number of possible ways to break such an exact symmetry by  introducing explicit breaking parameter in the various position of $ M_D $ \footnote{Note that authors in Ref.~\cite{Nath:2018hjx} have discussed the breaking of such symmetries by assigning breaking term in different places of $ M_D $ and $ M_R $. Moreover, we like to emphasize here that from the  model building prospect of view, one can consider this kind of explicit breaking term as the leading order correction terms in the original theory which then leads to non-zero leptogenesis together with phenomenologically viable neutrino mixing parameters (see Ref\cite{Karmakar:2014dva} where leading order corrections under $ A_4 $ symmetry has been discussed). }. Here, we give one example of such kind by assigning a breaking term  in the (3,1) position of  $ M_D $ which we renamed as $ M_D^{\prime}$ as mentioned in eq.~(\ref{eq:MdPrime}). 
%
%
Using eq.~(\ref{eq:MdPrime}), one finds the  mass matrix for the light neutrinos in the type-I seesaw formalism  as
\begin{eqnarray}\label{eq:MnuPrime}
 M^{\prime}_{\nu} & \simeq &  M_{\nu} -  \epsilon a ~ 
 \left( \begin{matrix}
0 &  0  &      i b r_2^{\prime}  \cr
0 & 0 &  ( a r_1^{\prime} + i b r_2^{\prime}) \cr
 i b r_2^{\prime} &  ( a r_1^{\prime} + i b r_2^{\prime}) & 2 ( a r_1^{\prime} - i b r_2^{\prime}) \cr
\end{matrix} \right) + \mathcal{O}(\epsilon^{2}), 
\end{eqnarray}
where $ M_{\nu} $ is defined in eq.(\ref{eq:tex_1}).

\begin{figure}[h!]
\hspace{-0.5cm}
\includegraphics[height=7cm,width=17cm]{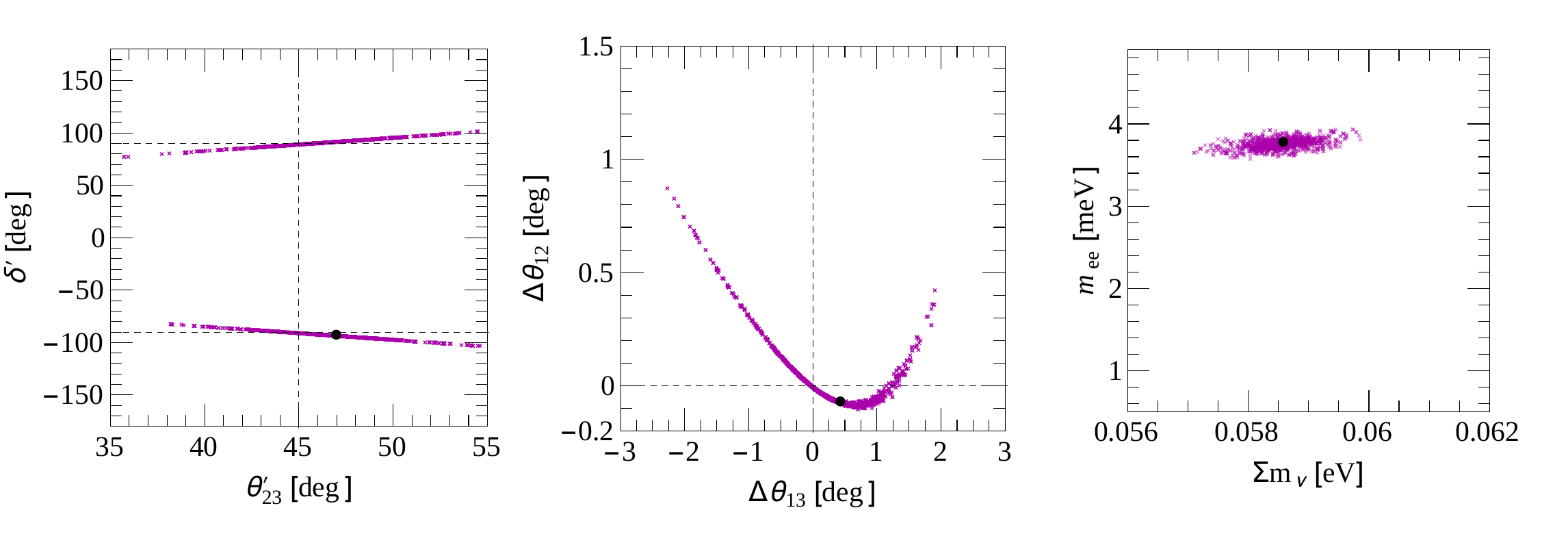}
\caption{\footnotesize Correlation between neutrino oscillation parameters. Here, all the magenta points have $ \chi^{2} < 30 $ whereas best-fit point  is denoted by black dot and which has minimum $ \chi^{2} $,  $ \chi^{2}_{\rm min} \simeq 0.88 $.}
\label{fig:NuOscParam_NH}
\end{figure}

Now to figure out the impact of such a breaking scheme on the neutrino masses and mixing angles, we diagonalize $M_\nu^\prime$ with the mixing matrix $V^\prime$, which has the same form as the mixing matrix $V$ (see eq.(\ref{eq:pmns})) in the limit $\epsilon \rightarrow 0$. To calculate the neutrino masses and mixing angles of eq.(\ref{eq:MnuPrime}), we consider three small parameters $\epsilon$, $\theta_{13}^{}$ and $\zeta = m_2^{}/m_3^{}$. We find neutrino masses as
\begin{align}
m_1^{\prime} = & ~ 0 \;, \nonumber \\
m_2^{\prime} \simeq   & ~ m_2 + \epsilon a c_{12} \sin\phi \left[2 c_{12} \chi_1 -  \sqrt{2} b r_2 ^{\prime} s_{12}   \right ]\;, \nonumber \\
 m_3^{\prime} \simeq  & ~  m_3 -   \epsilon ~ 2  a \cos\phi  ~\chi_2 \;,
\end{align}
whereas neutrino mixing angles are expressed as 
\begin{align}\label{eq:MixingAngleAfBr}
\theta_{12}^{\prime} \simeq & ~ \theta_{12} - \epsilon \dfrac{a \sin\phi \left[ \sqrt{2} b r_2 ^{\prime} \cos 2\theta_{12} + 2  \chi_1 \sin 2\theta_{12}\right] }{2 m_3 \zeta} \;, \nonumber \\
 \theta_{13}^{\prime} \simeq  & ~ \theta_{13} - \epsilon \frac{ a b r_2 ^{\prime} \sin\phi  }{\sqrt{2} m_3^{}} \;, \nonumber \\
 \theta_{23}^{\prime} \simeq  & ~\theta_{23} + \epsilon \dfrac{a \left[  a r_1 ^{\prime} \cos2\phi + b r_2 ^{\prime} \sin2 \phi \right]}{m_3} \;,
\end{align}
 where $ \chi_1 = (a r_1^{\prime} \sin\phi - b r_2^{\prime} \cos \phi ) $ and $ \chi_2 = (a r_1^{\prime} \cos\phi + b r_2^{\prime} \sin \phi ) $.
We now perform a detailed description of our numerical analysis and present them in fig.~(\ref{fig:NuOscParam_NH}) \footnote{Note that we also vary $ \epsilon \in [-1, 1] $ along with other parameters as mentioned  in eq.(\ref{eq:VaryParam}).}.

In first and second panel of fig.~(\ref{fig:NuOscParam_NH}), we show different correlations among neutrino oscillation parameters. On the other hand, we demonstrate our results  in ($ \sum m_{\nu} \times m_{ee} $)-plane in third panel, where $ \sum m_{\nu} $ represents the sum of absolute neutrino mass and  $ m_{ee}  $ signifies the effective Majorana neutrino mass matrix element \footnote{At the current juncture, the neutrinoless double beta-decay ($ 0\nu \beta \beta $) experiments  are the only experiments which can probe the Majorana nature of neutrinos and lepton number violation \cite{Pas:2015eia}. 
The effective Majorana neutrino mass matrix element $ m_{ee}  $  governs the rate of such process.}.
  In this figure, all the magenta points have $ \chi^{2} < 30 $ whereas best-fit scenario is shown by black dot which corresponds to minimum $ \chi^{2} $ ($ \chi^{2}_{\rm min} \simeq 0.88 $) eV. The unitary mixing matrix at $  \chi^{2}_{\rm min}  $ after symmetry breaking  is given by, 
\begin{eqnarray}
U = \left(
\begin{array}{ccc}
 0.8155 & 0.5596 & -0.0064+ i~ 0.1485 \\
 -0.3812 + i~ 0.0896 & -0.5605 + i~ 0.0613 & 0.6490 \\
 0.4163+ i~ 0.0836& -0.6009 + i~ 0.0572 & 0.7465 \\
\end{array}
\right) \;,
\end{eqnarray}
whereas the masses of three neutrinos are given as $ (m_1, m_2, m_3) = (0, 0.0086, 0.0488) $ eV. 

From  first plot, i.e., in ($ \delta^\prime, \theta^\prime_{23} $)-plane plot,  we notice two branches of solutions corresponding to maximal value of  Dirac CP-phase, $ \delta = \pm 90^\circ$.  A large deviation of around $ \sim 10^\circ $ for $ \theta_{23} $ from its maximal value i.e., $  \theta_{23} = 45^\circ $ has been observed. 
 For  $ \delta $, we also find a deviation of around $ \sim 10^\circ $ from its maximal value $ \pm 90^\circ $. Investigating second panel, we notice a negative as well as positive correlations between $ \Delta\theta_{13} $ and $ \Delta\theta_{12} $ around their $ \chi^{2}_{min} $. Also, one observes five times larger deviation in $ \Delta\theta_{13} $ compare to $ \Delta\theta_{12} $. Moreover, from last panel, 
we observe that  $\sum m_\nu^{} \sim 0.06~\mathrm{eV}$ which is expected to satisfy mass squared differences of current oscillation data for the vanishing lowest mass in case of NO. Also, the predicted $m_{ee}^{}$ is only around $ \sim $ 4 meV in this case. Such small values of $\sum m_\nu^{}$ and $m_{ee}^{}$ would be difficult to probe by $0\nu\beta\beta$ experiments and forthcoming cosmological observations respectively.
\section{Conclusion}\label{sec:Conclusion}
In this work, we  implement $ \mu - \tau $ reflection symmetry in the low energy neutrino mass matrix through type-I minimal seesaw formalism. This framework helps us to understand the theory behind the  origin of neutrino  masses and their flavor mixing. An immediate consequence of the concerned symmetry is that it leads to the atmospheric mixing angle,  $ \theta_{23} = \pi/4$ and the Dirac CP violating phase, $ \delta = \pm \pi/2$ together with  the Majorana phases, $ \rho, \sigma = 0, \pi/2 $.
It also predicts non-zero $ \theta_{13} $ unlike $ \mu - \tau $ permutation symmetry which predicts a vanishing $ \theta_{13} $. These predictions are in good agreement with the recent oscillation data. Whereas the minimal seesaw framework gives rise to a vanishing lightest neutrino mass (i.e., $ m_1 = 0 $) as this can still able to explain the observed two mass squared differences, namely solar ($ \Delta m^{2}_{21} $) and atmospheric ($ \Delta m^{2}_{31} $)
 mass squared differences.   Further, we endeavor to generate leptogenesis in the given framework. We find that the concerned model does not lead to a non-zero CP asymmetry unless one introduces a minimal perturbative term to the model. This term allows one to generate successful leptogenesis which is compatible with the latest experimentally observed result. Afterwards, we proceed to discuss various correlation study among high energy neutrino mass matrix elements as well as neutrino oscillation parameters, respectively.

 We give the analytical expressions of mixing angles, $ \theta_{13}, \theta_{12}$ in terms of model parameters along with the expressions of absolute neutrinos masses. Also, as  the concerned model is embroiled with a minimum number of free parameters which serves us to perform different analytical correlation studies among low and high energy parameters in the framework of $ \mu - \tau $ reflection symmetry. In this study, we find a good agreement between analytical study corresponding to their numerically analysis. Later, to find the matter-antimatter asymmetry of the  Universe, 
we observe vanishing CP asymmetry within the formalism of exact symmetry. To tackle the situation and to generate successful leptogenesis, we observe that a minimal perturbation of the model is able to generate experimentally observed leptogenesis. Further, we find that a parameter of $ \mathcal{O}(10^{-2}) $ is able to engender non-zero leptogenesis which further leads us to a sizable baryon-to-entropy ratio. This also break 
$ \mu - \tau $ reflection symmetry and leads to non-maximal  $ \theta_{23}, \delta $.
Moreover, some linear correlations among the parameters of high energy neutrino mass matrix elements have been observed.  We find that model parameter $ |a| $ lies in the parameter range (0.1 -- 0.8)$ v $ whereas $ |b| $ falls in the range (0.05 -- 0.35)$ v $. Similarly, both the parameters of $ M_R $ falls around $ \sim 10^{14} $ GeV to explain latest oscillation results. We also find that the lightest heavy Majorana neutrino mass, $ M_1 $ is around $ 10^{13} $ GeV in this model.
Among neutrino oscillation parameter, we notice a large deviation of $ \mathcal{O}(10^\circ) $ for  $ \theta_{23}$ from its maximal value while significant amount of deviation for  $\delta $ has been noted. Also, a five times larger deviation of $ \theta_{13} $ compared to $ \theta_{12} $ from their best-fit values has been observed. In the analysis, we also find that the given framework predicts a very small value of the effective Majorana neutrino mass matrix element, $ m_{ee} $ of around 4 meV. Also, the sum of the active neutrino masses, $ \sum m_{\nu} $ for NO of around 0.06 eV has been found in this scenario. 

Finally, we conclude this work with a note that given the status of latest global-fit of neutrino oscillation results, $ \mu - \tau $ reflection symmetry along with the minimal seesaw model comes out as one of the finest theoretical approach to understand the origin of neutrino mass and  flavor mixing. In this paper, we have made an attempt to address both these issues in  substantial details. However, more and more neutrino oscillation data from the ongoing and the forthcoming neutrino oscillation experiments, especially from those related to the measurement of the atmospheric mixing angle and the leptonic CP-violation phase,  will help to rule out or verify different flavor models and strengthen our theoretical understanding. 

\begin{acknowledgements}
Author is indebtedly thankful to  Prof. Shun Zhou  for many useful discussions, insightful comments and careful reading of the manuscript. Author also thanks Prof. Zhi-zhong Xing, Dr. Jue Zhang and Mr. Guo-yuan Huang for fruitful discussions. The research work of author was supported in part by the National Natural Science Foundation of China under grant No. 11775231. 
\end{acknowledgements}

\appendix
\section{$ M_{\nu} $ in scenario SII of $ M_D $ :}\label{sec:appendix}
The low energy neutrino mass neutrino matrix, $ M_{\nu} $ can also be constructed  considering SII texture of  $ M_D $ as given by eq.(\ref{eq:md_LSS}) (here we use $ n = 1, \eta = (2l+1)\pi, \phi_{m} = (2m+1)\pi $ with $m, n = 0,1,2...$) along with $ M_R $ as mentioned  by eq.(\ref{eq:mr}). Under type - I seesaw mechanism, one 
can express the final form of  $ M_{\nu} $ as
\begin{eqnarray}\label{eq:MnuCase2}
-M_{\nu} & = &  \dfrac{1}{{r_1^2-r_2^2}}  \left( \begin{array}{ccc}
 b^2 r_1 & -(b^2 r_1 + i a b r_2) & (b^2 r_1 - i a b r_2) \\
* & -\left(a^2-b^2\right) r_1 + 2 i a b r_2 & -\left(a^2+b^2\right)r_1 \\
* & * & -\left(a^2-b^2\right) r_1 - 2 i ab r_2 \\
\end{array}
\right) \;.
\end{eqnarray}
This can also be written as,
\begin{eqnarray}\label{eq:MnuAfterSeesaw2}
M_{\nu}   & = &  \left( \begin{matrix}A & B &  - B^{*} \cr
  * & C & D \cr
 * & * & C^{\ast} \cr
\end{matrix} \right), 
\end{eqnarray}
where, $ A, D $ are real and it has same form as eq.(\ref{eq:MnuAfterSeesaw}). Here, 
\begin{eqnarray}\label{eq:LowHighCase2}
A & = &  b^2 r^{\prime}_1 \;,  \nonumber \\
B & = & -(b^2 r^{\prime}_1 + i a b r^{\prime}_2)\;,  \nonumber \\
C & = & -\left(a^2-b^2\right) r^{\prime}_1 + 2 i a b r^{\prime}_2\;,  \nonumber \\
D & = & -\left(a^2+b^2\right)r^{\prime}_1 \;, 
\end{eqnarray}
with $ r^{\prime}_{1,2} = r_{1,2}/(r_1^2-r_2^2) $. Thus, we notice that this form of  $ M_{\nu} $ also possesses $ \mu - \tau $ reflection symmetry and gives  same predictions as mentioned in eq.(\ref{eq:prediction}). Note also that one gets similar form of low energy neutrino mass matrix either one chooses eq.(\ref{eq:md_LSS}), eq.(\ref{eq:mr}) instead of eq.(\ref{eq:MdHat}), eq.(\ref{eq:MrHat}) as they are related by 2-3 rotation matrix. Similarly, one can also show that the textures SIII, SIV lead to the same form of $ M_{\nu} $ and hence lead to $ \mu - \tau $ reflection symmetry.

\bibliography{mu-tau}
\end{document}